\documentclass[twocolumn]{aastex63}
\pdfoutput=1

\usepackage{amsmath}
\renewcommand{\thesubsection}{\Roman{subsection}}

\accepted{2020 September 24}
\submitjournal{ApJ}
\shorttitle{Hybrid simulations of parallel propagating Alfv\'en waves}
\shortauthors{Gonz\'alez et al. 2020}
\graphicspath{{./}{figures/}}

\begin{document}

\title{The role of parametric instabilities in turbulence generation and proton heating: Hybrid simulations of parallel propagating Alfv\'en waves}

\correspondingauthor{C.A. Gonz\'alez}
\email{carlos.gonzalez1@austin.utexas.edu}

\author{C.A. Gonz\'alez}
\affiliation{Department of Physics, The University of Texas at Austin, Austin, TX 78712, USA}
\author{A. Tenerani}
\affiliation{Department of Physics, The University of Texas at Austin, Austin, TX 78712, USA}
\author{M. Velli}
\affiliation{Department of Earth, Planetary, and Space Sciences, University of California, Los Angeles, CA, USA}
\author{P. Hellinger}
\affiliation{Astronomical Institute, CAS, Bocni II/1401, Prague CZ-14100, Czech Republic}
\affiliation{Institute of Atmospheric Physics, CAS, Bocni II/1401, Prague CZ-14100, Czech Republic}

\begin{abstract}
Large amplitude Alfv\'en waves tend to be unstable to parametric instabilities which result in a decay process of the initial wave into different daughter waves depending upon the amplitude of the fluctuations and the plasma beta. The propagation angle with respect to the mean magnetic field of the daughter waves plays an important role in determining the type of decay. In this paper, we revisit this problem by means of multi-dimensional hybrid simulations. In particular, we study the decay and the subsequent nonlinear evolution  of large-amplitude Alfv\'en waves by investigating the saturation mechanism of the instability and its final nonlinear state reached for different wave amplitudes and plasma beta conditions. As opposed to one-dimensional simulations where the Decay instability is suppressed for increasing plasma beta values, we find that the decay process in multi-dimensions persists at large values of the plasma beta via the filamentation/magnetosonic decay instabilities. In general, the decay process acts as a trigger both to develop a perpendicular turbulent cascade and to enhance mean field-aligned wave-particle interactions. We find indeed that the saturated state is characterized by a turbulent plasma displaying a field-aligned beam at the Alfv\'en speed and increased temperatures that we ascribe to the Landau resonance and pitch angle scattering in phase space.
\end{abstract}


\section{Introduction} \label{sec:intro}

Collisionless or weakly collisional turbulent plasmas are typically found in space and astrophysical environments. It is the case of the heliosphere and the solar wind, the outflow of plasma continually emitted by the sun and that permeates our solar system. Large amplitude fluctuations in the plasma velocity and magnetic field, known as Alfv\'enic fluctuations, are commonly observed in the solar wind. Such fluctuations are almost incompressible, and they display the typical velocity-magnetic field correlation that characterizes Alfv\'en waves propagating away from the sun \citep{ coleman1967,belcher1971}. In spite of such a high degree of correlation, Alfv\'enic fluctuations in the solar wind are characterized by a well-developed power-law spectrum that dominates the low-frequency range of the solar wind fluctuations energy \citep{bavassano, horbury_PPCF_2005, bruno2005}. It is thought that these Alfv\'enic fluctuations might be generated near the sun and that they may contribute to coronal heating and solar wind acceleration  \citep{Velli93,Erdelyi&Fedun_2007,verdini_ApJ_2009}, problems that are still under debate in the community. 

\textit{In-situ} observations  support the idea that dissipation of turbulent fluctuations might contribute significantly to plasma heating \citep{smith2001heating,bruno2005, hellinger_2013}. However, internal energy generation involves different channels, such as  resonant~\citep{Hollweg&Isenberg_2002,ChenEA2019} and non-resonant (stochastic) wave-particle interactions \citep{ CranmerEA2014, chen_2001, chandran_2010}, magnetic reconnection within coherent structures or scattering by  current sheets \citep{Dmitruk_2004,servidio2011,Zhdankin_2013,matthaeus2011,karimabadi2013,parashar2009,drake2009, Isliker2017,pisokas2018} among others, and the very nature of the dissipation process(es) is still puzzling. In this regard, the evolution of proton temperature in the solar wind shows a strong departure double adiabatic expansion, and preferential particle heating in the perpendicular direction to the local magnetic field is typically observed. Besides, the proton distribution function displays many non-thermal features such as a secondary proton population with a drift velocity of the order of the local Alfv\'en speed \citep{Marsch_2006}. Interestingly, kinetic simulations have shown that a field-aligned proton beam may form self-consistently through the decay of an initial large-amplitude Alfv\'enic fluctuation \citep{araneda2008proton,MatteiniEA2010}. 

Alfv\'en waves of arbitrary amplitude with constant total pressure are known to provide an exact solution to the compressible Magnetohydrodynamic system in a homogeneous plasma, in that nonlinearities are turned-off and there no couplings with compressible modes. However, such a dynamical system is linearly unstable to parametric instabilities and large amplitude Alfv\'en wave are known to decay into compressible and secondary Alfv\'enic modes through three or four-wave resonances that lead to a variety of parametric instabilities, depending on the plasma beta and dispersive effects. Such is the case of parametric decay \citep{galeev_sov_phys_1963,Derby_1978}, modulational, and beat instabilities \citep{sakai, wong,nariyuki,Jayanti&Hollweg_1993}. Parametric instabilities  of Alfv\'en waves (or of a spectrum of Alfv\'en waves) have been widely studied over the years through theoretical approaches \citep{Goldstein_1978,Jayanti&Hollweg_1993,malara_Phys_fluids_1996}, and numerical simulations adopting both MHD \citep{GhoshEA1994, Ghosh&Goldstein_1994,malara2, DelZannaEA2001} and kinetic models \citep{TerasawaEA1986,MatteiniEA2010,Verscharen2012,Nariyuki2012, Tenerani_2017} although most often in one dimensional setups.  In particular, the traditional parametric decay instability has attracted much attention over the years both in the context of turbulence and plasma heating. This type of decay is most efficient at low values of the plasma beta and it essentially involves the decay of a pump Alfv\'en wave into a lower frequency reflected Alfv\'en wave and a forward sound wave. For this reason, parametric decay remains an appealing process because it provides a natural mechanism for the production of reflected modes, which is essential for the triggering of a turbulent cascade. Indeed, recently it has  been proposed  as a viable mechanism to initiate the turbulent cascade in the solar wind acceleration region \citep{chandran_2018, reville_2018}, while global MHD simulations of the solar wind have also shown that the parametric decay instability can contribute substantially to solar wind heating and acceleration, thanks to the generation of compressible modes that, in the absence of kinetic effects,  naturally steepen into shocks (see, e.g., \citet{shoda}). The traditional parametric decay has been also invoked as a possible source for the generation of inward modes and solar wind turbulence in the inner heliosphere, where an increasing content of reflected waves (cross-helicity) and an evolving turbulent spectrum is observed with increasing heliocentric distance \citep{bavassano2000}. However,  expansion effects are known to inhibit its development,  essentially because the parametric decay process is strongly suppressed as the plasma beta increases at larger heliocentric distances \citep{anna1, tenerani_2020, DelZannaEA2015}. Temperature anisotropies  can destabilize the parametric decay at values of the plasma beta  approaching unity and above, but the anisotropy in the solar wind is not large enough to affect significantly the instability \citep{Tenerani_2017}.

Despite much work on parametric instabilities, less attention has been devoted to kinetic effects in multi-dimensions. The multidimensional nature of parametric instabilities of a parallel propagating Alfv\'en wave was first investigated via two-fluid linear theory by \citet{kuo1988} and later in the work by \citet{vinas1991a,vinas1991b} where it was shown that the oblique propagation of the daughter waves allows for additional parametric instabilities depending on the angle of the density perturbation with respect to the mean magnetic field. Previous numerical studies showed that while oblique modes naturally emerge when the pump wave itself is in oblique propagation or in two-dimensional turbulence (as observed for example in \cite{MatteiniEA2010, primavera_2019}), perpendicular and quasi perpendicular modes can grow as the result of a different decay process of an Alfv\'en wave in parallel propagation, known as the filamentation and the magneto-acoustic instability, respectively. Such highly oblique modes have been reported previously in  numerical simulations \citep{GaoEA2013,comicsel2018,comicsel2019}.

In this paper we revisit the stability of  Alfv\'en waves in parallel propagation using 1D, 2D and 3D hybrid simulations to explore the combined effect of multi-dimensionality and kinetic proton physics at different values of the plasma beta and pump wave amplitude. We consider left-handed circularly polarized large-amplitude Alfv\'en waves and we  investigate how the decay process and its saturation and nonlinear stages depend on the pump wave amplitude, plasma beta, and dimensionality.  We find that the overall decay process involves a superposition of modes due to parametric and filamentation instability that survives at values of the plasma beta well above unity.  The instability triggers both a turbulent cascade in the perpendicular direction and a wave energy conversion process that ultimately leads to the formation of a field-aligned proton beam at the Alfv\'en speed, regardless of the plasma beta, and that appears to be associated with a strong particle heating. While particle heating is affected by the pump wave amplitude, it is surprisingly observed whenever the decay occurs, regardless of the dimensionality and of the plasma beta. This result  thus suggests that the particle heating we observe in our simulations is predominantly a one-dimensional process driven by the decay of the wave.

This paper is organized as follows: In section \ref{model} we present the quasi-neutral hybrid model and the numerical setup that we have employed in this study. Section \ref{results} is divided in three parts. In subsection \ref{global_d} we describe the global dynamics of the instability for different plasma beta, wave amplitude, and dimensionality of the problem. The spectral properties of the electromagnetic field are presented in subsection \ref{spectral_prop} where we focus on the turbulent properties at different scales. Finally, in subsection \ref{proton_heat} we discuss the effect of the proton/electron beta and the wave amplitude on the proton heating and acceleration, and address the problem of wave-particle interactions by proposing a possible mechanism to explain the observed features. In section \ref{discussion} we summarize our results.

\section{Model and simulation setup}
\label{model}

We have employed a hybrid model where electrons are treated as a massless and isothermal neutralizing fluid while the proton dynamics is described by the Vlasov-Maxwell equations (Eqs. \ref{Eq1}). The coupling with the electromagnetic fields is given by the low-frequency and non-relativistic Maxwell's equations, where quasi-neutrality ($n_i=n_e = n$) is assumed and the electric field is determined via the generalized Ohm's law:

\begin{subequations}
\begin{equation}
\frac{ \partial f_i } {\partial t} + \textbf{v} \cdot \frac{\partial f_i}{\partial \textbf{r}} + \frac{e}{m_i}\left( \textbf{E} +  \frac{\textbf{v}}{c} \times \textbf{B} \right) \cdot \frac{\partial f_i}{\partial \textbf{v}} = 0
\label{eq:New1} 
\end{equation}
\begin{equation}
\frac{\partial \textbf{B}}{\partial t} =  - c \  \nabla \times \textbf{E}, \ \ \ \ \ \textbf{J} = \frac{c}{4 \pi} \nabla \times \textbf{B}
\label{eq:Max1}
\end{equation}
\begin{equation}
\textbf{E} + \frac{\textbf{u}_i}{c} \times \textbf{B} =   - \frac{k_B T_e \nabla n }{ e n}  + \frac{ \textbf{J} \times \textbf{B}}{ e n}  + \eta \nabla \times \textbf{B},
\label{ohm}
\end{equation}
\label{Eq1}
\end{subequations}

\begin{deluxetable}{c c c c c c c c c c c}
\tablenum{1}
\tablecaption{Initial conditions for the simulations presented in this paper. \label{table1}}
\tablewidth{0pt}
\tablehead{\colhead{Run}  & \colhead{$\Delta x$} & \colhead{$\Delta t$} & \colhead{ppc} & \colhead{$\beta_p$} & \colhead{$\beta_e$}  & \colhead{$\delta b_0$} & \colhead{$\eta$}}
\startdata
A1-1D  & 0.25 & 0.025 & 10000 & 0.25  & 0.25  & 1.0 & 0.002\\
A2-1D & 0.25 & 0.025 & 10000 & 0.5 & 0.5 & 1.0 &  0.002\\
A3-1D & 0.25 & 0.025 & 10000 & 2.0  & 2.0  &  1.0  & 0.002 \\
B-1D  & 0.25 & 0.025 & 10000 & 0.5 & 0.5 & 0.5   & 0.002\\
B-1D & 0.25 & 0.025 & 10000 & 0.5 & 0.5 &  0.25   & 0.002 \\
A1-2D & 0.0625 & 0.005 & 1000 & 0.25 & 0.25 & 1.0 & 0.0004\\
A2-2D & 0.0625 & 0.00625 & 1000 & 0.5 & 0.5 & 1.0  & 0.0004\\
A3-2D &  0.0625 & 0.01 & 1000 & 2.0 & 2.0 & 1.0 & 0.0004\\
B1-2D & 0.25 & 0.025 & 1024 & 0.5 & 0.5 & 1.0  & 0.002\\\ 
B2-2D  & 0.25 & 0.025 & 1024 & 0.75 & 0.75 & 1.0  & 0.002\\
B3-2D & 0.25 & 0.025 & 1024 & 1.0 & 1.0 & 1.0   & 0.002\\
B4-2D &  0.25 & 0.025 & 1024 & 2.0 & 2.0 & 1.0  & 0.002\\
B5-2D & 0.25 & 0.025 & 1024 & 4.0 & 4.0 & 1.0  & 0.002\\
C1-2D  & 0.25 & 0.025 & 1024 & 0.5  & 0.5 & 0.5 & 0.002\\
C2-2D & 0.25 & 0.025 & 1024 & 0.5 & 0.5 &  0.25 & 0.002\\
D-2D & 0.25 & 0.025 & 1024 & 0.0 & 2.0 & 1.0  & 0.002\\
A3-3D & 0.5 & 0.05 & 512 &  2.0 & 2.0 &  1.0 & 0.004
\enddata
\end{deluxetable}

 with $c$ the speed of light, $e$ the electron charge, $k_B$ the Boltzmann constant and $T_e$ is the electron temperature. The proton number density $n$ and the proton bulk velocity $\textbf{u}_i$ are computed from the moments of the distribution function ($n = \int{f(\textbf{r},\textbf{v},t) d\textbf{v} }$ and $n \textbf{u}_i = \int{ \textbf{v} f(\textbf{r},\textbf{v},t) d\textbf{v}}$ respectively). In this work we made use of the CAMELIA code (see e.g.  \citet{Franci_2018}), which is a hybrid particle-in-cell code that uses the current advance method \citep{matthews1994} and Boris scheme for the particle pusher, with good stability and long term accuracy.
 
The numerical setup consists of a large amplitude, large scale Alfv\'en wave propagating along the mean magnetic field ${\bf B}_0$, that we take  along the \textit{$x$-axis}. Periodic boundary conditions are imposed in all directions of the computational box. Lengths are normalized to the proton inertial length $d_i =  c/\omega_{p}$ with $\omega_p = (4\pi ne^2/m_i)^{1/2}$ the proton plasma frequency. Time is expressed in units of the inverse of proton gyrofrequency $\Omega_{ci}^{-1} = (eB_0/m_i c)^{-1}$ and velocities are normalized to the Alfv\'en speed $v_A = B_0/(4 \pi n m_i)^{1/2}$.  The plasma beta for both ions and electrons is defined as $\beta_{p,e}=8 \pi n k_B T_{p,e}/B_0^2$. Since in almost all simulations we have $\beta_p=\beta_e$, we will just use the symbol $\beta$ to indicate the proton beta, unless otherwise specified. We have included the resistive term in the generalized Ohm's law to improve energy conservation by avoiding energy accumulation at the grid scales. The resistive coefficient is defined in units of $4 \pi \omega_{p}^{-1}$ and the associated length scale is chosen to be greater than the grid size but smaller than any other scale of interest (i.e. smaller than the proton inertial length or proton gyroradius depending on the plasma beta).
 
We initialize the system using an isotropic homogeneous plasma with uniform particle density and proton velocities randomly distributed with a Maxwellian distribution function at given temperature $T_p$ and a fixed number of particles per cell (npp). The initial pump Alfv\'en wave is initialized with a wave number $n_0=4$ and wave vector $k_0 = 2\pi n_0/L$, $L$ being the  box  size  in  units  of  proton  inertial  length  (we  use a square or cube box with equal sides), that satisfies the condition $\delta {\bf u} = -(\omega_0/k_0) \delta {\bf b}$, with $|\delta {\bf b}|=\delta b_0$ the amplitude of the pump wave normalized to the mean magnetic field magnitude $B_0$.  The wave frequency is determined from the normalized dispersion relation $k_0^2=\omega_0^2/(1-\omega_0)$ for left-handed circularly polarized waves. The initial Alfv\'en wave is given by $\delta b_z = \delta b_0 \cos{(k_0 x)}$ and $ \delta b_y = - \delta b_0 \sin{(k_o x)}$. The box size adopted in all the simulations presented along this paper have $L=128 d_i$ and the pump wave is weakly dispersive with a wavenumber $k_0 v_A/\Omega_{ci} = 0.196$. A summary of the numerical and plasma parameters adopted in this work can be found in Table~\ref{table1}.

\section{Results}
\label{results}

\subsection{Global dynamics}
\label{global_d}

\begin{figure}
\includegraphics[width=0.45\textwidth]{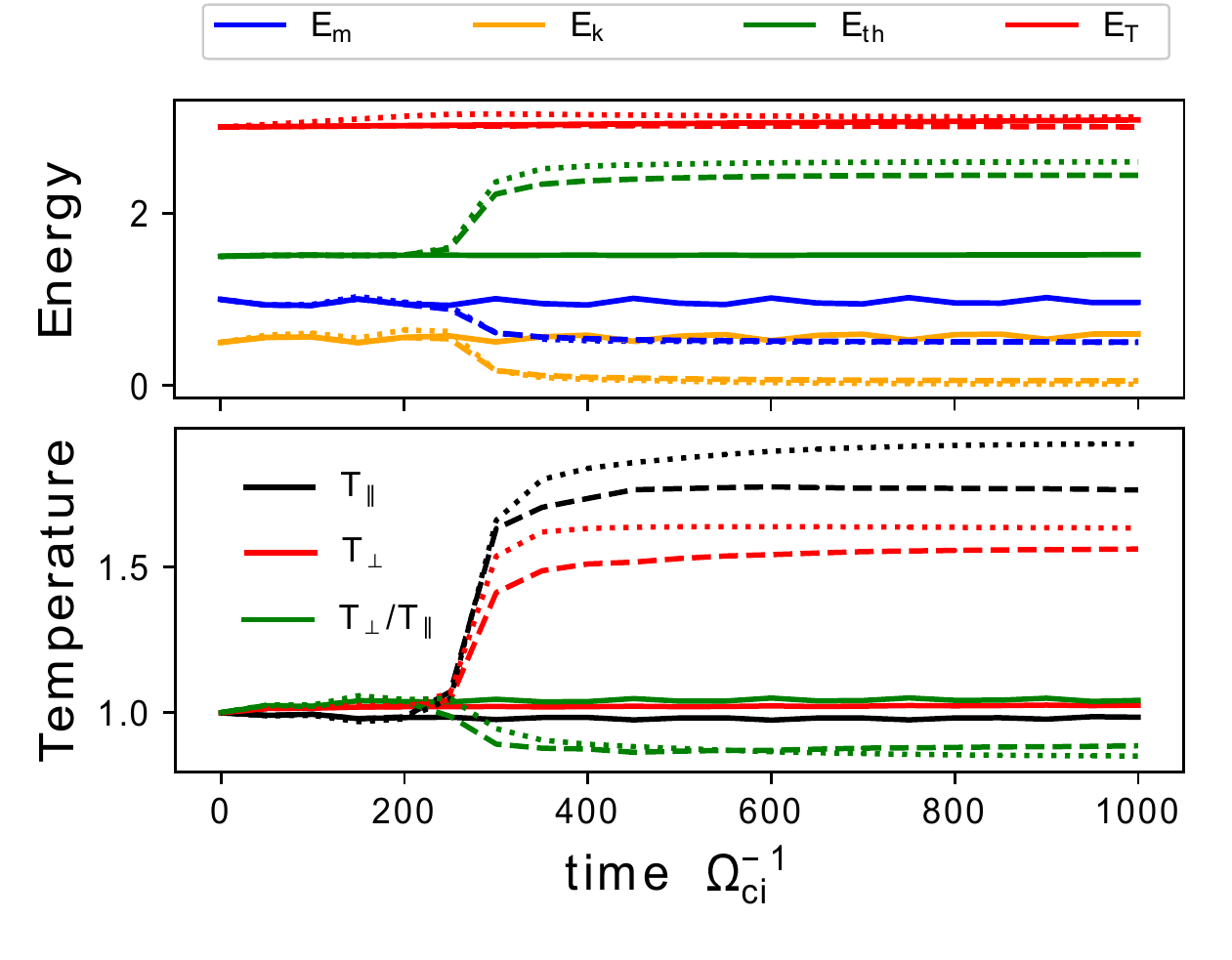}
\caption{(\textbf {Top.}) Temporal evolution of the energy and the proton temperature for run A3-1D (solid lines),  A3-2D (dashed lines) and  A3-3D (dotted lines)). Magnetic (blue), kinetic energy (orange), thermal energy (green)  and total energy (red). (\textbf{Bottom}). Parallel temperature (black), perpendicular temperature (red) and the ratio between parallel and perpendicular components (green)}
\label{Fig1}
\end{figure}

In Fig.~\ref{Fig1}, top panel,  we show the time evolution of the kinetic, magnetic and thermal energy of three representative simulations that we use as a reference to summarize the main properties of the dynamical evolution of the system. In particular, we show results for runs A3-1D, A3-2D and A3-3D for the 1D, 2D, and 3D case, respectively, with $\beta=2$ and $\delta b_0=1$. The bottom panel shows the parallel and perpendicular proton temperature evolution (black and red colors, respectively) and the temperature anisotropy $T_\bot/T_\parallel$ (green color) for the same simulations. We have introduced the parallel and perpendicular temperatures defined in terms of the decomposition of the pressure tensor according to the direction of the total magnetic field as: $p_\parallel = \textbf{p} \colon \hat{\textbf{b}}\hat{\textbf{b}}$ and  $p_\perp = \textbf{p} \colon (\mathbb{\textbf{I}} - \hat{\textbf{b}}\hat{\textbf{b}})/2$. The pressure tensor $\textbf{p} = \int(u-v_p)_i(u-v_p)_j f({\bf r,v},t) d{\bf v}$ is obtained from the particle velocity distribution and $\hat{\textbf{b}} = \textbf{B}/\Vert \textbf{B} \Vert$ is the direction of the total magnetic field. 

\begin{figure*}
\includegraphics[width=\textwidth]{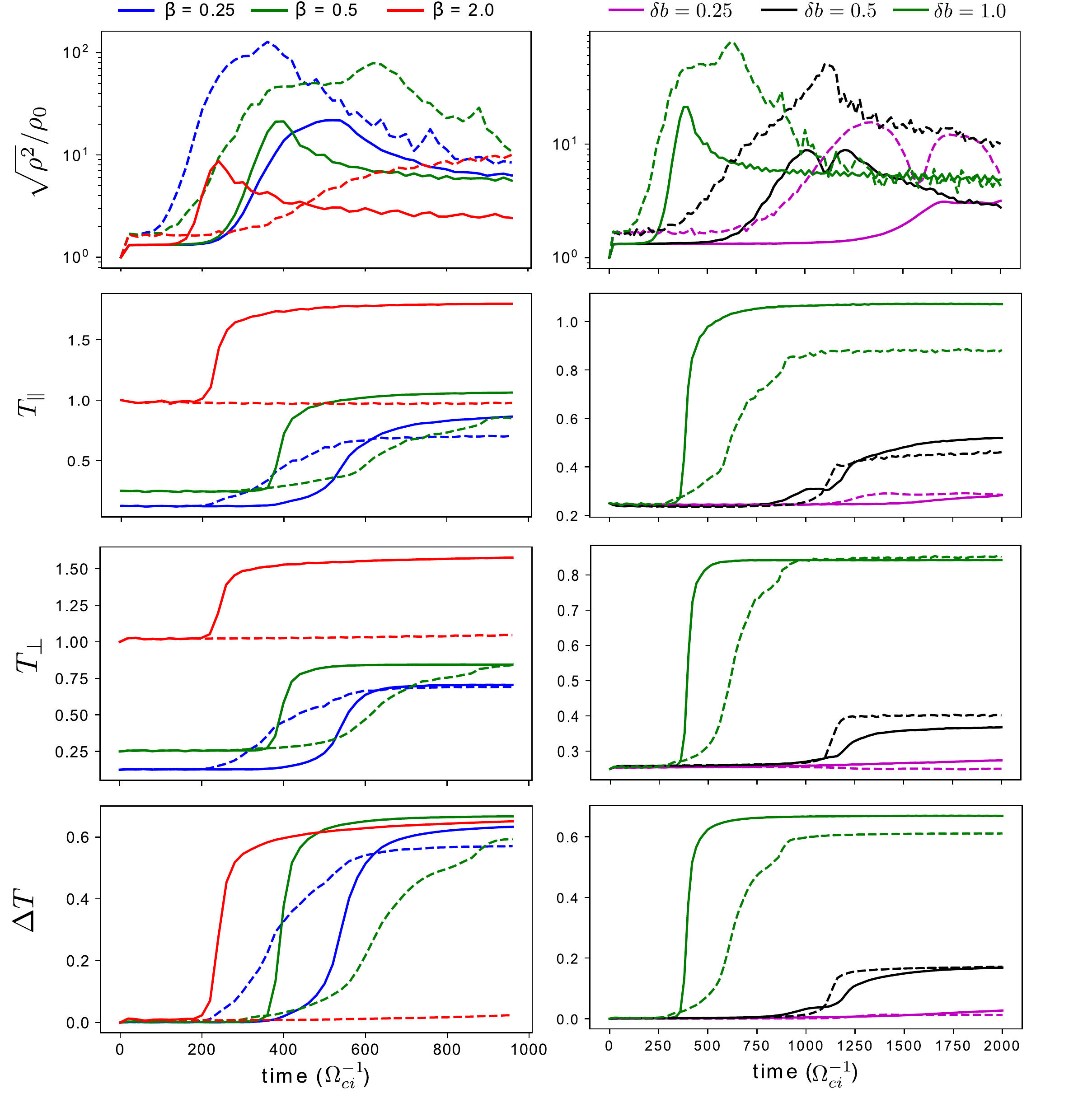}
\caption{The rms of the density, mean parallel temperature, mean perpendicular temperature and 
temperature difference in time for 1D simulations (dashed lines) and 2D simulations (solid lines).}
\label{Fig2}
\end{figure*}

Three different stages can be identified during the evolution: initially, the wave propagates without significant dispersion, the kinetic and magnetic energy oscillate around a mean value while the proton temperature remains constant. This oscillation is possibly due to fact that the initial condition is not an exact solution to the Vlasov-Maxwell equation \citep{sonnerup1967large}.After this initial stage, the pump wave decays  by conveying its energy to the particles (at $t \sim 250\Omega_{ci}$), resulting in an increase of the overall thermal energy. Finally, the saturation of the instability slows down the particle energization process, and the system achieves a steady state condition with almost constant kinetic, magnetic and thermal energy ($t\sim 400\Omega_{ci}$). Note that for the 1D simulation shown here there is not proton heating at all. That is because for $\beta=2$ the decay instability is suppressed in 1D, and therefore the wave is not disrupted. The total energy of the system is not perfectly conserved during the simulations, in fact, there is a relative error of the order of $3 \%$ and some numerical heating is present in the simulations.

The evolution of the root mean square (rms) of density fluctuations and of the average proton temperatures for different plasma beta and wave amplitudes are presented in Fig~\ref{Fig2}, where the  decay of the pump wave is marked by the rapid increase of density fluctuations and of the overall temperature. Here, the total temperature is defined as $T=(T_\parallel + 2T_\perp)/3$ and $\Delta T = T(t)-T(t=0)$ represents the net change of the total temperature from the initial value. Results for 1D simulations are also plotted as a reference  (dashed lines).

The left panels of Fig.~\ref{Fig2} display results for different plasma beta for an initial wave amplitude $\delta b_0=1$. As can be seen, the evolution of the decay process  is consistent with the predictions from Hall-MHD linear theory in the 1D cases. The pump wave is subject to decay instability which becomes slower as the (electron) plasma beta increases. For the amplitudes considered here, the instability is suppressed at large plasma beta ($\beta=2$), where the wave is more likely to decay via modulational or beat instabilities, with smaller growth rates and hence a slower decay process. The slight increase in density fluctuations that can be seen in the upper panel of Fig.~\ref{Fig2} for the 1D case does not indeed correspond to the disruption of the pump Alfv\'en wave.  Interestingly, the 2D simulations display an opposite trend with the plasma beta. In the low beta regime the 2D simulations are characterized by a growth rate similar to the 1D cases, although the decay occurs later than in the 1D, in agreement with previous studies comparing parametric decay from one to three dimensions \citep{DelZannaEA2001}. However, the 2D simulations display a rapid decay process even in the large beta case and, contrary to the 1D case, the growth rate tends to increase with the plasma beta.  We ascribe such differences between the 2D and 1D set of simulations to the onset of filamentation/magnetosonic decay simultaneously to the main parametric decay process, and to the resulting nonlinear dynamics. Besides, a large increase of thermal energy is always observed when the decay occurs. Although multi-dimensional simulations display a slightly larger final temperature than the 1D cases, this fact points to the fact that most of the heating mechanism(s) may be ascribed to a 1D dynamics. We defer a discussion of proton heating to Sec.~\ref{proton_heat}.

\begin{figure}
\includegraphics[width=0.45\textwidth]{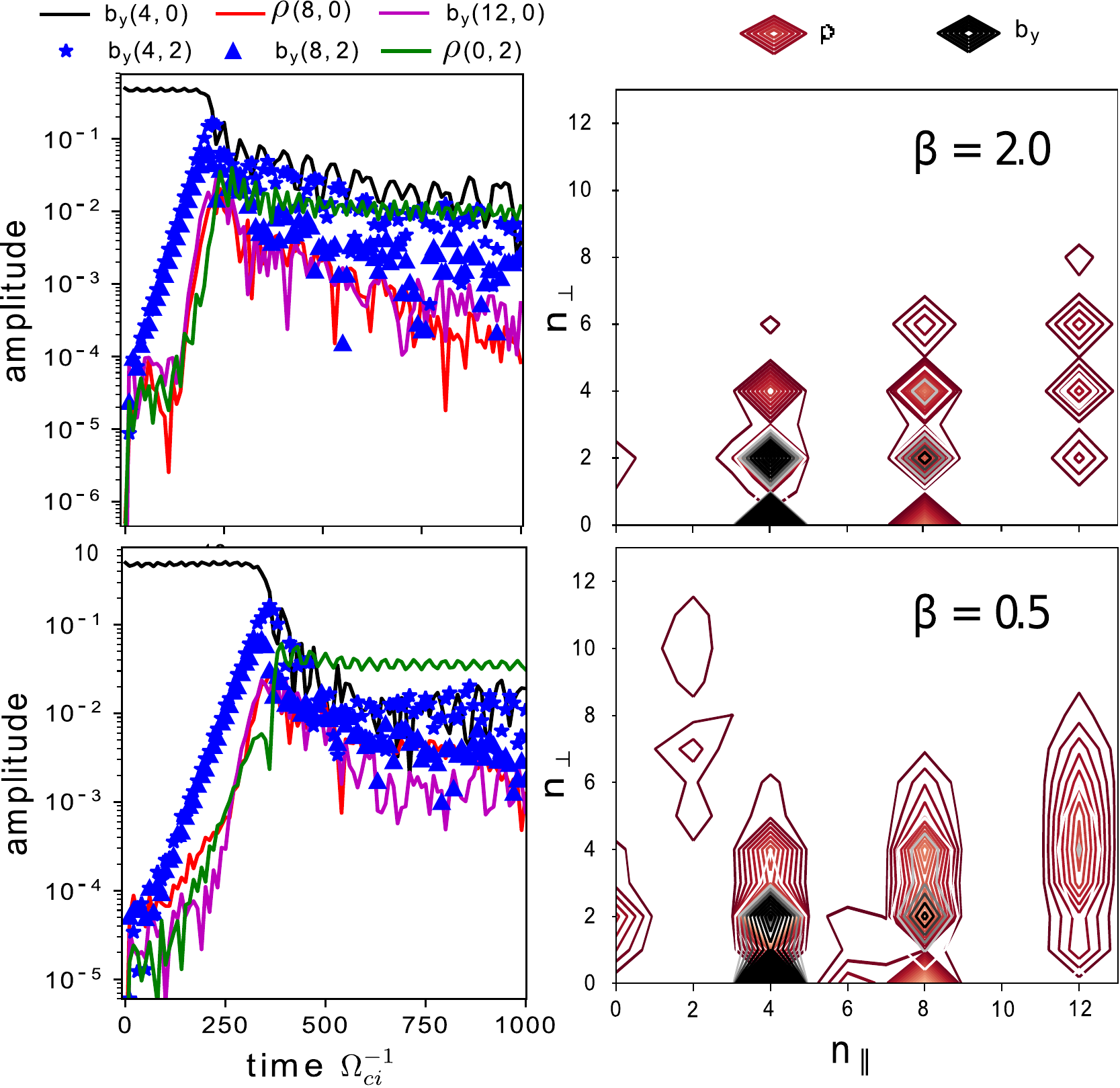}
\caption{(\textbf{Left}). Temporal evolution of the amplitudes of the most most unstable modes and of the pump wave (black) in runs A3-2D (\textbf{Top})  and A2-2D (\textbf{Bottom}).(\textbf{Right}). Superposition of 2D Fourier spectra of $B_y$ (black contours) and density $\rho$ (red contours) at the maximum of the the most unstable modes shown on the right panels.}
\label{Fig3}
\end{figure}

The right panels of Fig.~\ref{Fig2} show the same quantities on the left panel but for fixed beta ($\beta=0.5$) at different initial wave amplitudes. As can be seen by inspection, the growth rate increases with the amplitude, as is expected from linear theory.  Interestingly enough, a strong proton heating is observed as the pump wave amplitude increases, with  parallel and perpendicular temperatures displaying the same trends for 1D and 2D simulations. 

\begin{figure}
\includegraphics[width=0.45\textwidth]{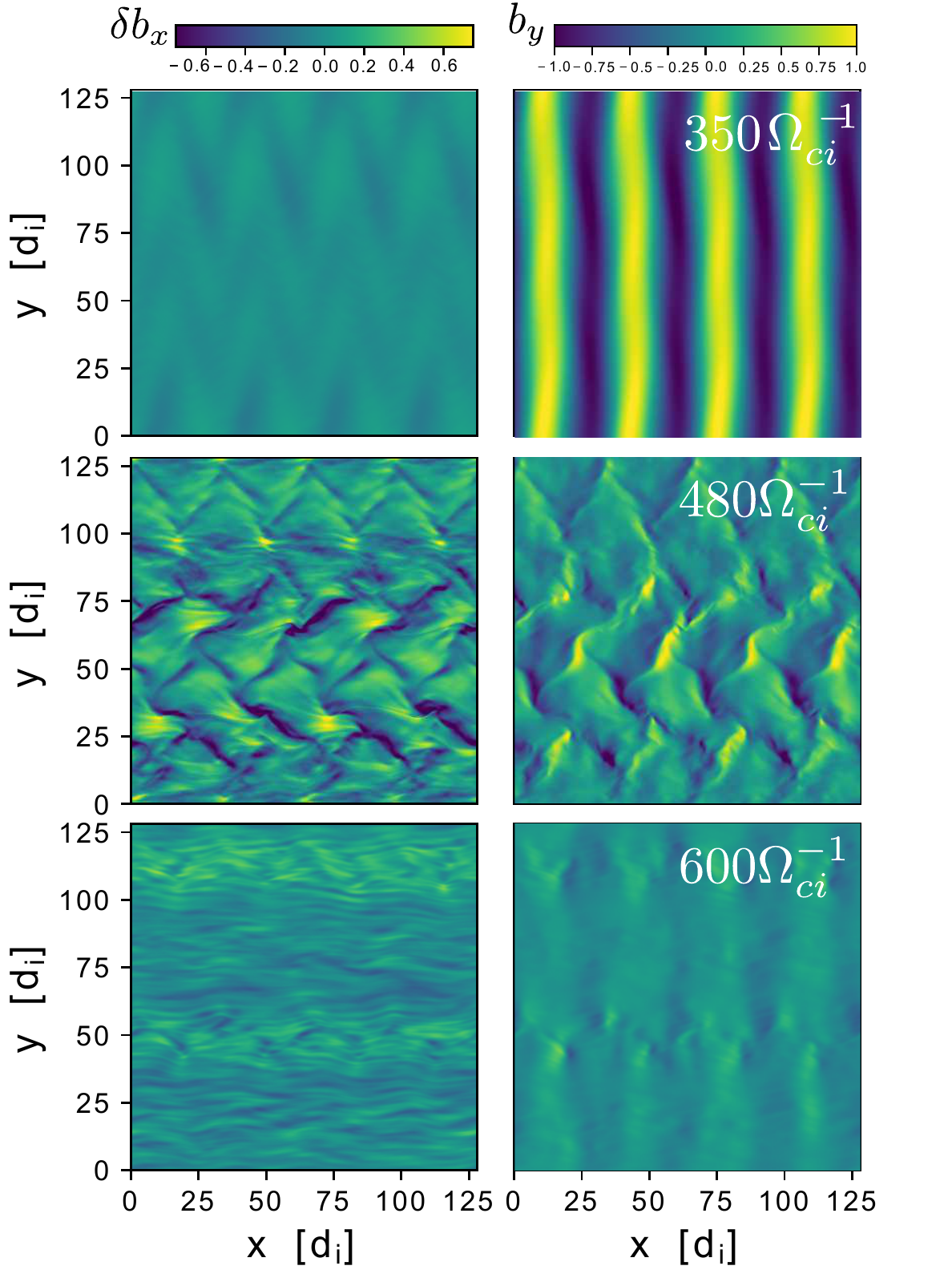}
\caption{2D Contour plot of the fluctuations of field-aligned component $\delta b_x$ (left column) and $b_y$ component of the magnetic field (right column) during three different stages of the evolution for run A2-2D.}
\label{Fig4}
\end{figure}

By way of illustration, the decay process for A2-2D and A3-2D is presented in Fig.~\ref{Fig3}. On the left panels we plot the amplitude of the most unstable modes and of the pump wave (in black). The right  panels show the superposition of the 2D FFT of $\rho$ (red contours) and $B_y$ (black contours)  at the maximum of the curves in the left panels.

\begin{figure*}
\includegraphics[width=\textwidth]{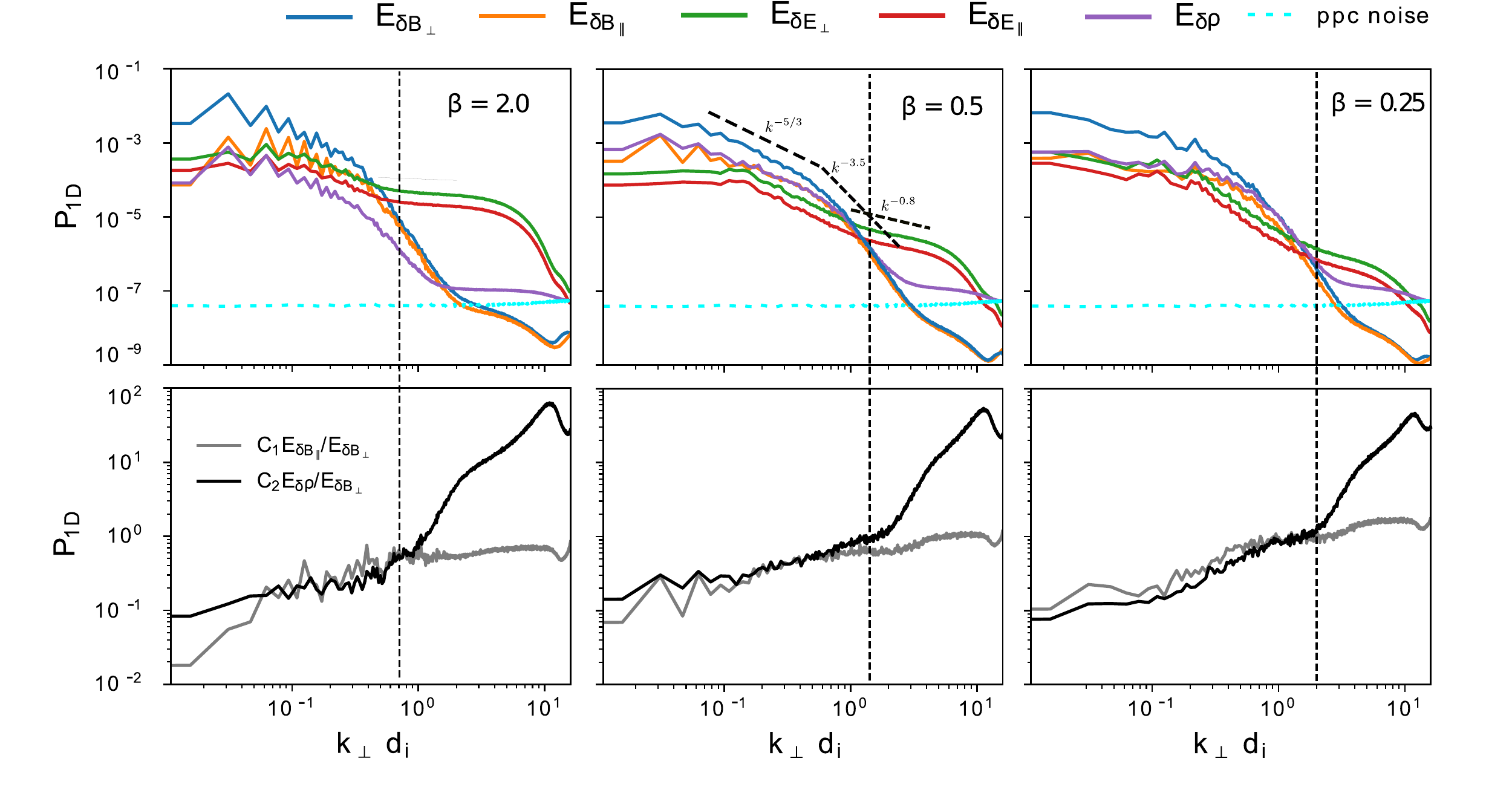}
\caption{(\textbf{Top}). The reduced 1D energy spectra as function of $k_\perp$ for $B_\perp$ (blue), $B_\parallel$ (orange), $E_\perp$ (green) and $E_\parallel$ (red) at the saturation stage for each simulation. The reduced spectrum for the initial proton density is plotted in cyan color as a reference of the particle noise level. (\textbf{Bottom}). Spectral ratio between $B_\parallel$ (gray) and $\rho$ (black) with $B_\perp$ normalized with the KAW linear prediction for each simulation.}
\label{Fig5}
\end{figure*}

As can be seen, different kinds of daughter waves are excited which lead to a competition between different types of parametric instabilities. We find that two types of decay are at play: a parallel and quasi-parallel one, corresponding to the traditional parametric decay instability, and a perpendicular one, corresponding to the filamentation/magnetosonic instability. The parallel decay is evident in the enhancement of density fluctuations with wave number $(n_\parallel,n_\perp)=(8,0)$ and, correspondingly, of the forward propagating Alfv\'en wave with $(n_\parallel,n_\perp)=(12,0)$, in agreement with the three-wave coupling resonance condition. The quasi-parallel side-band modes are also observed for $B_y$ and $\rho$ with wave number $(8,2)$ and $(4,2)$ (density is not shown), respectively. These quasi-parallel modes are the most unstable ones in the simulations with $\beta=2$ and $\beta=0.5$ and the maximum amplitude of each daughter wave corresponds to about $10 \% $ of the amplitude of the pump wave. The oblique daughter wave leads to an enhancement of $\rho$ and $B_y$ at wave number $(0,2)$. This mode is non-propagating and it is weakly damped, its amplitude remaining constant  throughout the simulation after the onset of filamentation/magneto-sonic instability.

For the sake of completeness we  display in Fig.~\ref{Fig4} the contour plots of the field-aligned component of the fluctuating magnetic field $\delta b_x$ (left panels) and of the pump wave  $b_y$ (right panels) at three different time for run A2-2D. The magnetic fluctuations of $ \delta b_x$ is found to be highly anti-correlated with density perturbations (not shown), a signature of the slow mode character of the growing fluctuations that persist even after the posterior distortion of the pump wave. The combination of the pump wave with the daughter waves and also the dispersion generated by small scales fluctuations leads to the steepening of the waveform (see middle panel of Fig.~\ref{Fig4}) that finally results in the disruption of the wave and the corresponding proton heating. The quasi-perpendicular mode $(0,2)$ can be easily identified in the contour of $b_x$ component since the amplitude of that mode remains constant after the saturation of the instability, contrary to the parallel and quasi-parallel modes that are highly damped after the saturation of the instability. 

\subsection{Spectral properties}
\label{spectral_prop}

The decay of the parent Alfv\'en wave into secondary modes triggers nonlinear interactions that ultimately lead to the establishment of a turbulent cascade. At saturation of the instability, an energy spectrum spanning scales down to sub-proton scales develops preferentially in the perpendicular direction to the mean magnetic field. In Fig.~\ref{Fig5} we show the resulting magnetic and electric field energy spectra for the 2D cases shown in the left panels of Fig.~\ref{Fig2}. In the top panels we plot the reduced 1D perpendicular spectrum of the parallel ($E_{B_\parallel}$, $E_{E_\parallel}$) and perpendicular ($E_{B_\bot}$, $E_{E_\bot}$) components of the electromagnetic field fluctuations at saturation stage. We also plot the reduced spectrum of density fluctuations for the same period and also the initial density spectrum as a reference to quantify the noise floor level in the simulations. The reduced 1D perpendicular spectrum is computed as $E_{\delta A}(k_\bot) = \int{ d k_\parallel} A(k_\parallel,k_\perp)$, with $A(k_\parallel,k_\perp)$ the 2D spectral energy density. The vertical dashed line marks the location of $k_\bot\rho_i=1$, with $\rho_i= \sqrt{\beta_i} d_i$ the proton gyroradius.

The power spectrum of the magnetic field components in the parallel direction is less developed and the spectrum is dominated by the daughter waves and its harmonics (not shown), but in the transverse direction the magnetic field shows a broad inertial range with a Kolmogorov-like spectrum ($\sim k_\bot^{-5/3}$). A spectral break is also observed when approaching to proton scales, marking the transition to another turbulent regime at sub-proton scales.  The spectral break occurs at the larger of the proton scales depending on the plasma beta, in agreement with previous numerical simulations \citep{franci_2016}. Turbulence at kinetic scales does not seem  to follow a universal behaviour, unlike the low-frequency, large scale dynamics.  A strong variability of the spectral index at sub-proton scales has indeed been reported in the solar wind and earth's magnetosphere with values ranging between -4 and -2 \citep{AlexandrovaEA2009,SahraouiEA2010,chen2016recent,Bowen2020}. Kinetic Alfv\'en waves (KAW) are often invoked to explain turbulent fluctuations at sub-proton scales, since magnetic field energy spectra measured by \textit{in-situ} observations and numerical simulation of plasma turbulence usually find power-laws with a spectral index close to -7/3, in agreement with KAW  theory. In this particular set of simulations, the spectrum at sub-proton scales is steeper than the KAW predictions and the spectral index is about -3.5. 

\begin{figure}
\includegraphics[width=0.45\textwidth]{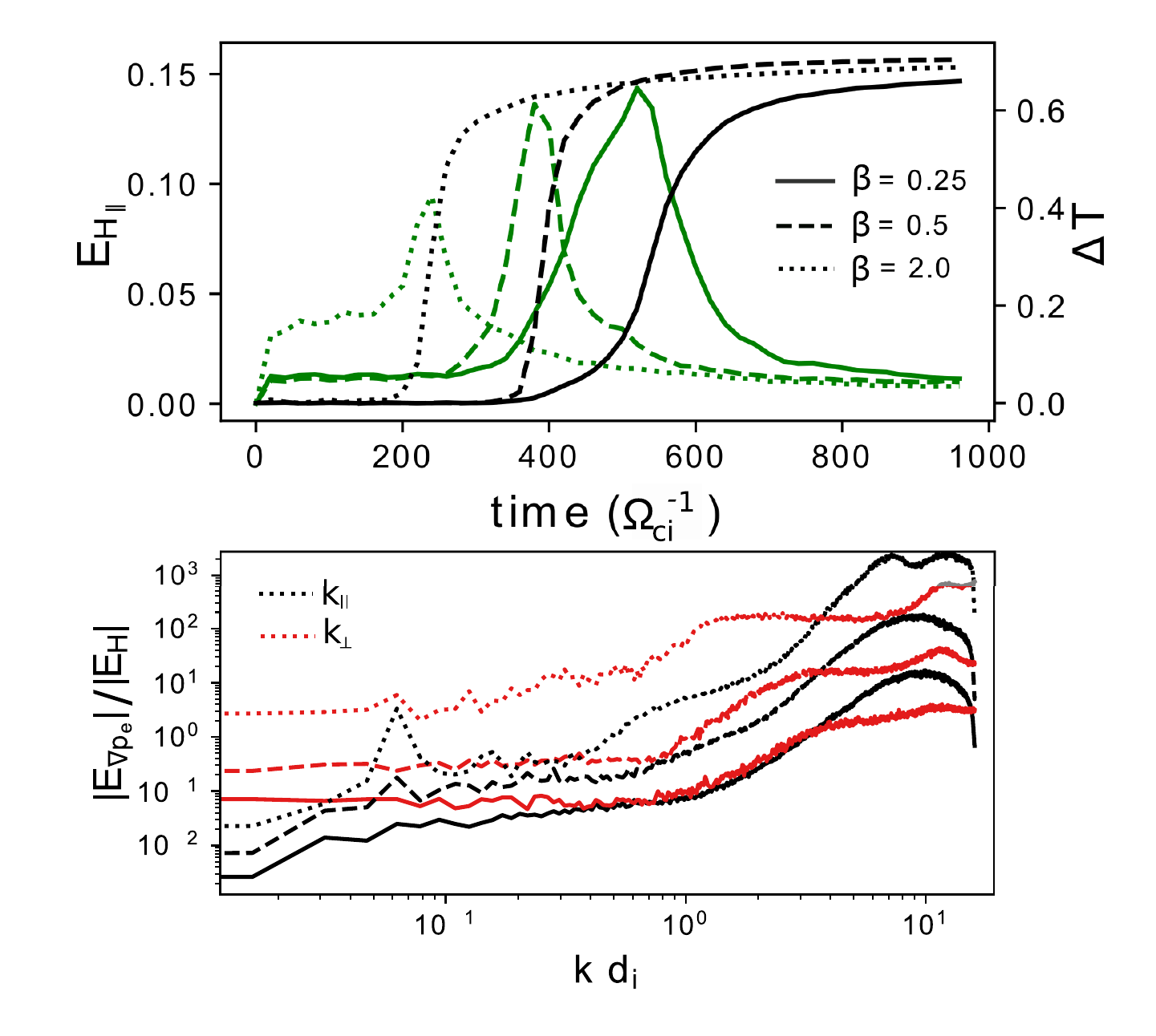}
\caption{(\textbf{Top}). Temporal evolution of the rms of the parallel Hall electric field component  (green lines) and the total temperature change for different plasma beta simulations. (\textbf{Bottom}).  Spectral ratio of the parallel component of electron pressure to the parallel Hall term for the same simulations shown in the bottom panel.}
\label{Fig6}
\end{figure}

The change of turbulence regime from the large to the small scales is marked by the increase of plasma compressibility, an effect that we observe in all of our sets of simulations. In order to characterize small scale fluctuations, we consider spectral field ratios that are known to provide a useful tool to investigate the polarization properties of turbulent fluctuations \citep{gary2009short,chen2013nature,grovselj2017,chen2017nature,cerri2019kinetic}. In the bottom panels of Fig.~\ref{Fig5} we present the spectral ratio  $R_1\equiv C_1 E_{ \delta B_\parallel}/E_{\delta B_\perp}$ and $R_2\equiv C_2 E_{\delta \rho}/E_{\delta B_\perp}$ for each simulation. The ratios are normalized to the rms of each field and to the theoretical prediction from KAW at different plasma beta ($C_1 =  \beta_t(1 + T_e/T_i) \ / \ 2 + \beta_t(1 + T_e/T_i)$ and $ C_2 = 4 \ / \ ((1 + T_i/T_e)(2 + \beta_t(1 + T_e/T_i)))$ with $\beta_t = \beta_p + \beta_e$. KAW theory predicts a value of unity for $R_1$ and $R_2$ at sub-proton scales. This would correspond to strong compressive magnetic fluctuations in nearly pressure balance in the kinetic range. In our set of simulations, where we considered $T_e/T_i=1 $, the values of $R_{1,2}$ for the magnetic field are in rough agreement with the theory, but the level of density fluctuations at scales smaller than proton scales largely exceeds the linear prediction. This could be due to nonlinearities and/or the presence of different wave activity like slow modes, whistler or others plasma modes in the sub-proton range. It is also noted that particle noise beyond proton scales may also dominate the density spectrum and therefore the ratio overestimates the expected values. 

The electric field shows a flattening of the spectrum at sub-proton scales with an index of about -0.8 at kinetic scales. This feature is observed in all simulations for different plasma beta. This correspond to the increment of propagation velocity of the turbulence fluctuations at sub-proton scales and it is well recognized to be due to the dominance of Hall term at proton scales which has been previously discussed in the context of fluid and kinetic simulations \citep{dmitruk2006structure,howes2011gyrokinetic,franci2015solar}.

\begin{figure}
\includegraphics[width=0.45\textwidth]{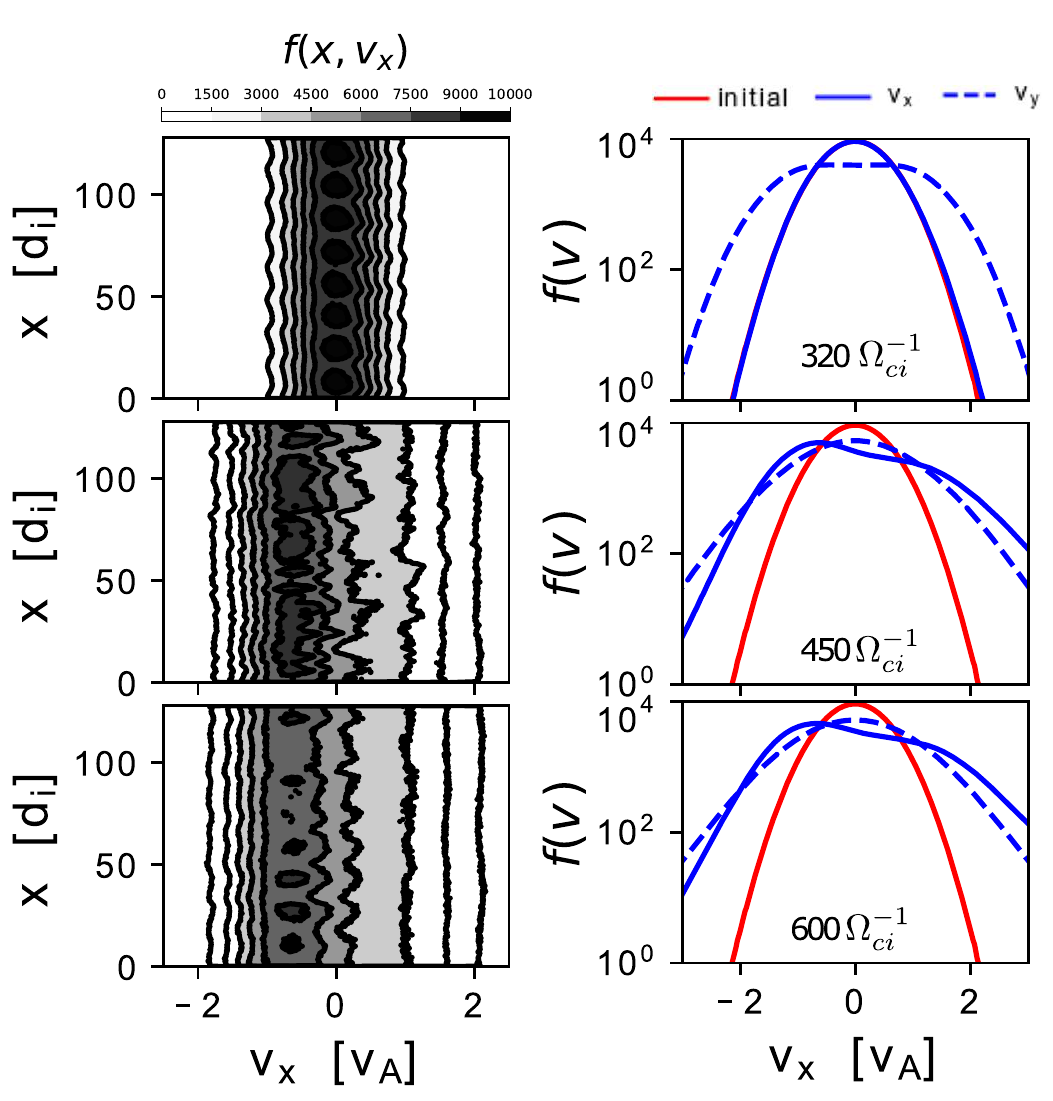}
\caption{(\textbf{Left}). Reduced distribution functions $f(x,v_x)$ at different stages of the evolution for the run A2-2D.  (\textbf{Right}). Reduced distribution function $f(v_x)$ for the x-component of the particle velocities (solid blue lines) and y-component of the particle velocities (dashed lines). The initial reduced distribution function is plotted in red}
\label{Fig7}
\end{figure}

The parallel electric field is developed during the collapsing stage of the pump wave and it plays a crucial role in the particle heating observed during the decay process until saturation. Interestingly, an important/dominant contribution to the parallel electric field comes from the field-aligned component of the Hall term rather than from the electron pressure term. Indeed, the electron pressure presents the same trend as the Hall term, but with smaller amplitude (not shown). In the top panel of Fig.~\ref{Fig6}, we show the rms value of the parallel component of the Hall electric field ($E_{H_\parallel} = E_{H_x}$) and $\Delta T$ for different beta simulations. As can be seen, there is a clear correlation between the Hall parallel electric field and particle heating in this set of simulations. In the bottom panel of Fig.~\ref{Fig6} we present the parallel (black lines) and perpendicular (red lines) spectral ratio between those two terms. Solid, dashed and dotted lines refer to the same plasma beta cases reported in the top panel, with each curve being taken at the maximum of the parallel electric field and averaged over $50 \Omega_{ci}$.  

At large scales, the ratio between electron pressure to Hall term at transverse scales ($k_\bot$) appears to increase as the plasma beta increases. This trend does not come as a surprise since  $P_{1D}(E_{\nabla p_e})/P_{1D}(E_{H})\propto\beta_e^2/4$ and it is consistent with the formation of slow modes along the perpendicular direction. In the parallel direction ($k_\parallel$), instead, the Hall term is larger than the electron pressure term at all beta values, as a consequence of the presence of strong electron currents produced by the decay of the pump wave. However, at sub-proton scales, the electron pressure dominates over the Hall term in both parallel and perpendicular directions. Again, particle noise may contribute to overestimate the level of density fluctuations observed at small scales. 

As already mentioned, the generation of parallel electric field fluctuations is crucial for the particle heating and acceleration observed during the decay process of the pump wave and until the steady-state condition is reached at saturation stage, a problem that we address in the next section.

\subsection{Proton heating}
\label{proton_heat}

The proton dynamics is illustrated in Fig.~\ref{Fig7} for the simulation with $\beta=0.5$ and $\delta b_0=1$. The left panels show contours of the proton distribution function in phase space ($x-v_x$) at three different times from top to bottom. The corresponding reduced distribution functions of parallel and perpendicular velocities ($v_x$ and $v_y$) are shown on the right panels. We show the particle information at three different stages of the evolution.  At $t=320\Omega_{c}^{-1}$, phase space vortexes form with the same wavelength of the density fluctuations developed by the parametric instability, with not yet significant heating at that time of the evolution. After this initial stage, a ``piston-like'' mechanism mediated by the parallel electric field allows for the generation of a secondary proton population propagating parallel to the mean magnetic field. The beam travels at the Alfv\'en speed, with signatures of particle trapping clearly visible in the phase space. The proton beam is persistent and remains stable when a steady state condition is reached.

\begin{figure}
\includegraphics[width=0.45\textwidth]{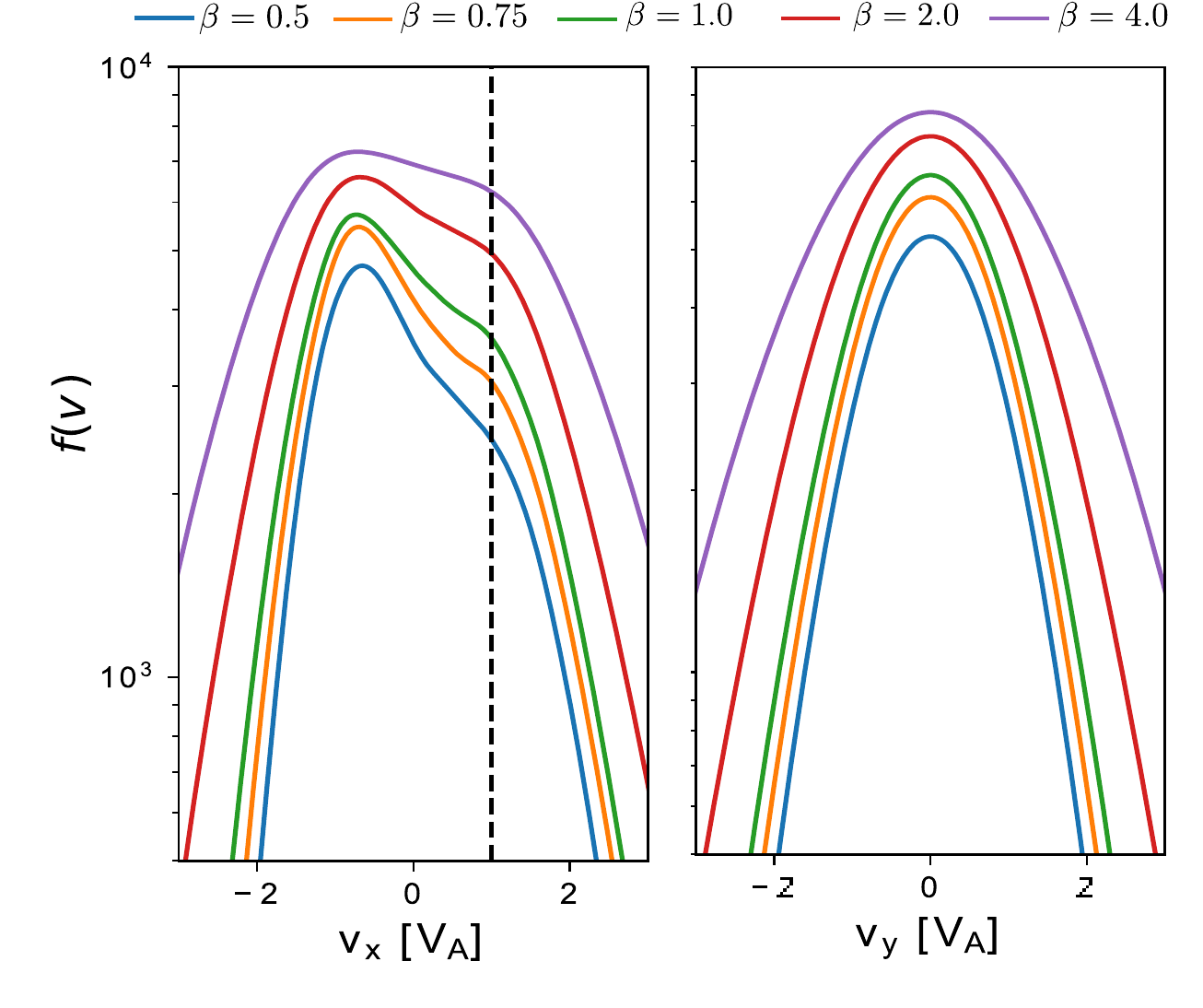}
\caption{(\textbf{Left}). Reduced distribution functions $f(v_x)$ and (\textbf{Right}) for $f(v_y)$
averaged over the steady state stage for run B1-2D (blue), B2-2D
(orange), B3-2D (green), B4-2D (red) and B5-2D (purple).}
\label{Fig9}
\end{figure}

The averaged particle distribution function (PDF) over $100  \Omega_{ci}$ after saturation stage is presented in Fig.~\ref{Fig9} for simulations with different plasma beta. We computed the average PDF right after the end of the instability is established and when the particle temperature is statistically constant. It can be noted that the beam forms around the Alfv\'en speed for all the beta cases. The distribution function in the perpendicular direction, instead, is a Maxwellian and the total change of perpendicular temperature is not affected by the plasma beta.

\begin{figure}
\includegraphics[width=0.45\textwidth]{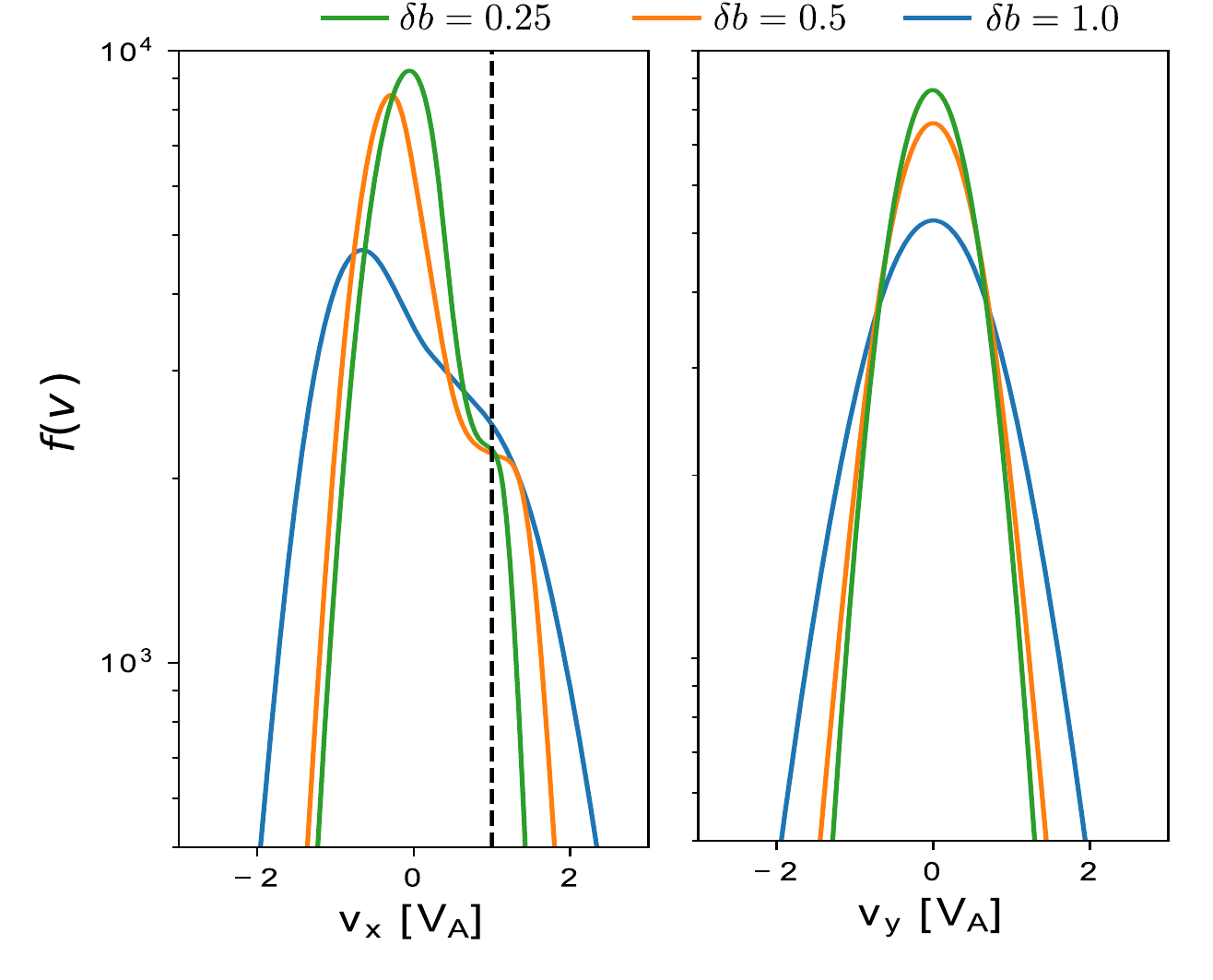}
\caption{(\textbf{Left}). Reduced distribution functions $f(v_x)$ and (\textbf{Right}) for $f(v_y)$
averaged over the steady state stage for runs B1-2D (blue), C1-2D(orange) and C2-2D (green).}
\label{Fig10}
\end{figure}

The effect of the wave amplitude on the final distribution function is presented in Fig. \ref{Fig10}. Even if the heating of particles depends on the amplitude of the pump wave, the beam formation along the mean magnetic field is persistent. The core of the distribution is mostly affected by the finite amplitude effects and larger tails in the distribution are found with larger wave amplitude. Since the distribution function is averaged over several gyroperiods, the number of particles in the beam is affected by the averaging.

The understanding of the overall proton heating due to the unstable behavior of Alfv\'en waves and the corresponding energy transport toward smaller scales is fundamental to understand the implications on the plasma heating typically observed in solar and astrophysical context. The nature of parallel and perpendicular heating comes from different physical mechanisms and, therefore, we first focus on the parallel heating, which is in fact due mainly to the beam generation, and subsequently we discuss the perpendicular heating and the possible mechanism.

\begin{figure}
\includegraphics[width=0.45\textwidth]{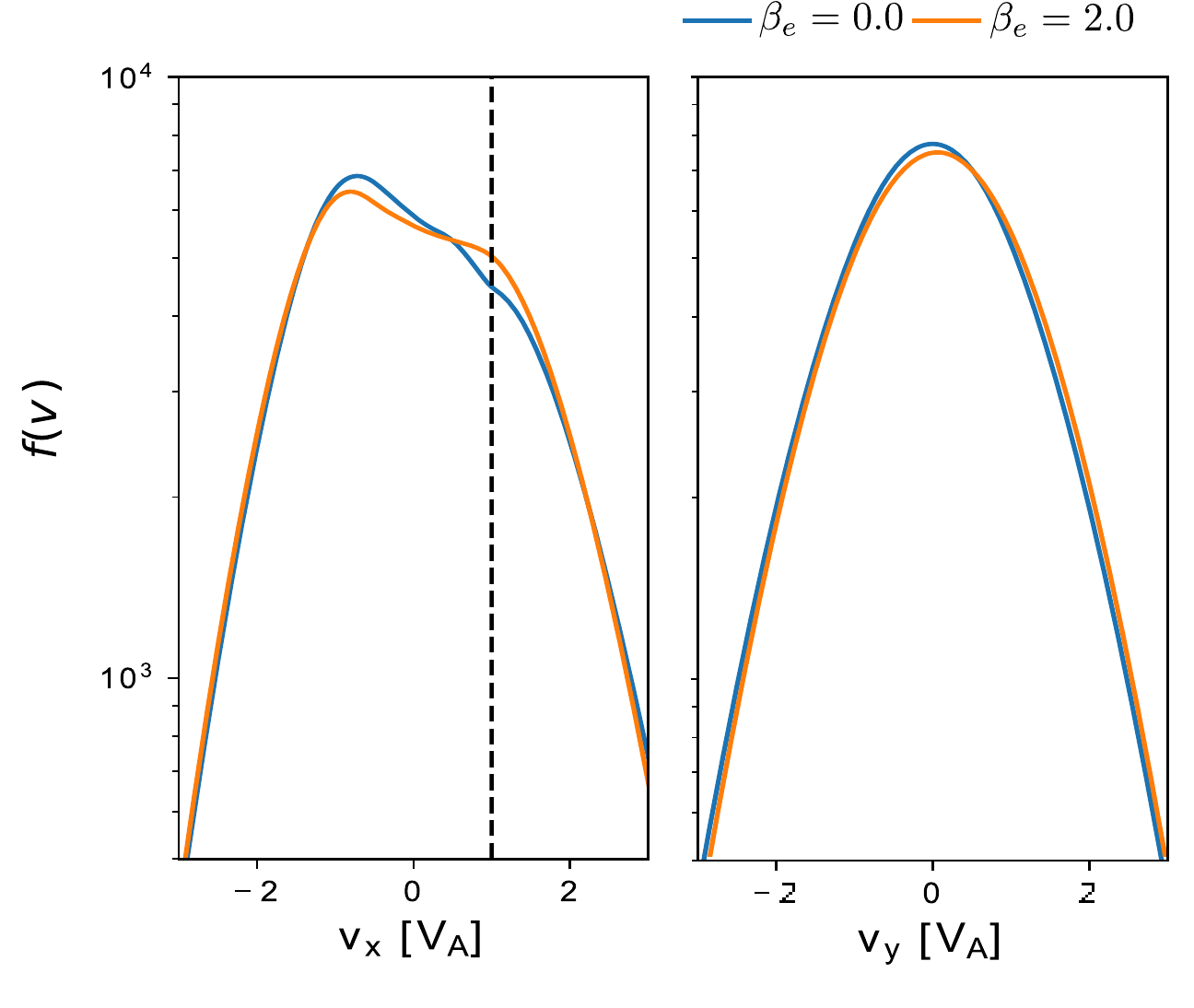}
\caption{(\textbf{Left}). Reduced distribution functions $f(v_x)$ and (\textbf{Right}) for $f(v_y)$
averaged over the steady state stage for runs D-2D (blue) and B1-2D (orange).}
\label{Fig11}
\end{figure}

According to previous work, the electron beta plays a crucial role on the saturation process because the electron temperature contributes to the strength of the parallel electric field via the pressure gradient term in the Ohm's law and also provide the coupling with the acoustic mode (see Eq.(\ref{eq:Max1})). It was suggested that the saturation of the instability was due to particle trapping by the field-aligned electric field generated by density fluctuations, which would lead to a mean field-aligned beam whose velocity appears to depend on the plasma beta. Hybrid simulations with $\beta_e=0$ have shown that the beam formation is suppressed and the saturation process results on the steepening of the ion acoustic wave, just like in the fluid description \citep{MatteiniEA2010}. In the same spirit, we present in Fig. \ref{Fig11} the results for simulations with $\beta_p=0.5$ and  $\delta b_0=1.0$ for two different scenarios: a case with ($\beta_e = 0.5$) and a case with cold electrons ($\beta_e = 0$).

As can be seen, the two simulations do not display significant differences in the PDFs.  Not only the perpendicular temperature achieved at the end of the process is the same for both simulations (right panel) but, importantly,  the field-aligned beam (left panel) is persistent in a plasma with $\beta_e=0$, even though a less populated beam is observed for cold electrons. This is because the electron pressure gradient still contributes to the trapping of  particles. The comparison between these two setups therefore points to the fact that it is the Hall term that contributes the most to the field-aligned electric field and to the beam formation.

According to Vlasov linear theory, proton heating by Alfv\'en waves is possible via resonant damping and particles can exchange energy with the wave only at discrete resonance interaction restricted by the condition $\omega - k_\parallel v_\parallel  = n \Omega_{ci}$,  with $\omega$ the frequency of the wave, $v_\parallel$ is the particle velocity and $k_\parallel$ is the wave number parallel to the magnetic field. Linear theory implies that for $n=0$, when the phase speed of the wave is of the order of the proton parallel velocity, particles can resonate with the wave and then they can gain or lose parallel velocity ($v_\parallel \leq \omega/ k_\parallel$ or $v_\parallel \geq \omega/ k_\parallel$ respectively). This resonant wave-particle interaction at $n=0$ represents two different physical interaction,  the Landau damping, driven by a parallel electric field, and  the transit time damping (TTD) \citep{fisk1976acceleration,achterberg1981nature}, which is the magnetic analog of Landau damping. In TTD it is the mirror force ${\bf F}_{mir}=\mu \boldsymbol\nabla_\parallel {B}$, with $\mu = m v_\bot^2/2B$ the conserved particle magnetic moment, to play the analogous role of the parallel electric field in Landau damping and, as such, it can also contribute to the parallel heating. Interactions with $n \neq 0$ instead lead to cyclotron resonances in which the particles resonate with the oscillating electric and magnetic field~\citep{Hollweg&Isenberg_2002}. This kind of wave-particle interaction results in the violation of the conservation of $\mu$ so that particles can experience strong perpendicular energization. In general these wave-particle resonances are important because they can lead to parallel/perpendicular heating produced by pitch-angle scattering resulting in an isotropization process of the particle distribution function by particle diffusion in velocity space \citep{kennel1966velocity,lynn2012resonance,Lynn_2013}. 

\begin{figure}
\includegraphics[width=0.45\textwidth]{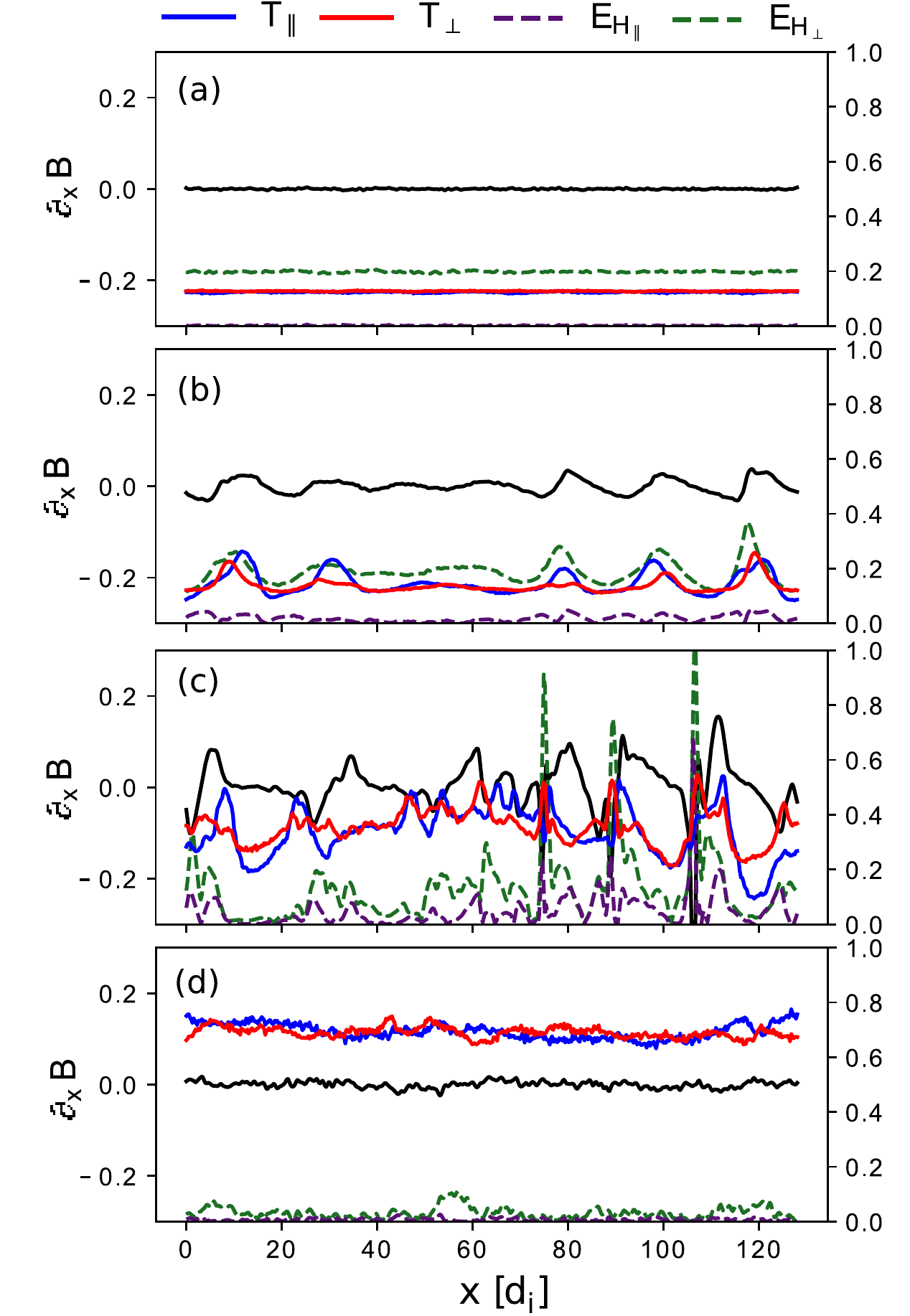}
\caption{Spatial profiles of different quantities for run A1D-I
at four different stages of the evolution: (a) at $t=100 \Omega_{ci}^{-1}$, (b) at $t=200 \Omega_{ci}^{-1} $, (c) at $t=400 \Omega_{ci}^{-1} $ and  (c) at $t=700 \Omega_{ci}^{-1}$. (black) the gradients of the magnitude $B$,
the (blue) parallel and (red) perpendicular proton
temperatures, (purple) parallel (green) and perpendicular  components of the Hall electric field. The y-axis scale on the right hand side is used for
the proton temperatures and the electric fields.}
\label{Fig12}
\end{figure}

As we have shown in section \ref{global_d}, the proton heating does not display major differences between 1D and 2D simulations whenever the wave is subject to parametric instabilities. In fact, the perpendicular heating is roughly the same, although a somewhat larger parallel temperature is obtained in the 2D simulations (see for instance left panel of Fig 2). In this sense, the main proton heating mechanism at play seems to be the same process regardless  of the dimensionality of the system and of the proton/electron plasma beta. In this way we can analyze the proton heating process for a 1D case without loss of generality. Moreover, the fact that in these numerical experiments the proton heating is essentially  a one-dimensional mechanism, we can rule out contributions from stochastic heating possibly due to KAWs or in general to obliquely propagating modes, and reconnection, which cannot develop in our 1D system.   

In Figure \ref{Fig12}, we present the spatial profiles of the parallel and perpendicular temperature and also the Hall term components ($E_{H_\parallel} = b_z j_y - b_y j_z$ and $E_{H_\perp} = \sqrt{E_{H_y}^2 + E_{H_z}^2} = B_0\sqrt{j_z^2 + j_y^2}$) and the gradients of the magnetic field ($\partial_x {B} = - {B}^{-1/2} E_{H_\parallel}$ with the current component $j_y = -\partial_x b_z$ and $j_z = \partial_x b_y$, at different stages of the evolution for run A1D-I. Initially, the circularly polarized Alfv\'en wave is characterized by a constant-${B}$ state and zero parallel Hall electric field because of the initial condition (Fig.~\ref{Fig12} (a)).  Once the decay comes into play, the disruption of the wave together with dispersive effects generate gradients of ${B}$ in the field-aligned direction and the wave tends to steepen in some localized regions.  There it follows an enhancement not only of density fluctuations but also of the Hall electric field through the current density components $j_y$ and $j_z$. Such steepened wavefronts with enhanced Hall electric field propagate at the Alfv\'en speed (not shown). The combined effect of particle trapping by the growing electric field fluctuations, and acceleration from the localized Hall electric field at the steepened fronts of the Alfv\'en wave contribute to the acceleration of particles into a field-aligned beam at the Alfv\'en speed, enhancing the number of resonating particles with the fluctuations themselves. We suggest that both type of $n=0$ resonances (Landau and TTD damping associated to the gradients of $B$) might be responsible for the parallel proton heating (Fig\ref{Fig12} (b-c)). Once the resonating fluctuation energy is transformed into particle heating, the system achieves a steady state, with small gradients of $B$, a persistent beam and a strong parallel and perpendicular heating (Fig\ref{Fig12} (d)). 

As mentioned above, perpendicular heating might be produced by pith-angle scattering processes due to the acceleration and redistribution of particles in the field-aligned direction of phase space.  There is also a high correlation between large perpendicular temperature and strong perpendicular Hall electric field  (Fig.~\ref{Fig12} (b-c)). However, we notice that the increment of parallel and perpendicular temperatures occurs simultaneously and at the same rate, favoring the idea that perpendicular heating is probably due to pitch-angle scattering. While here we have borrowed concepts from linear theory, the formation of the beam and the subsequent wave-particle interactions leading to saturation and plasma heating are really a nonlinear process where finite amplitude effects on particle orbits should be taken into account. We plan to develop a more detailed investigation of the beam formation and its possible relation to both wave steepening and perpendicular heating in future work.

It is important to also mention that the interaction of protons with current sheets and fluctuations, that develop naturally as a result of the turbulent cascade,  may account for the additional particle heating that we observed in the 2D and 3D simulations (e.g.~\cite{Dmitruk_2004, Isliker2017, pisokas2018, Zhdankin_2013}). However, we observe a more efficient parallel than perpendicular heating (cf.  Fig.~\ref{Fig2}, second and third panels), contrary  to what is expected from the strong turbulence perspective,  where the effects of current sheets results in strong perpendicular proton heating.

\section{Summary and conclusion}
\label{discussion}

We have performed hybrid simulations of large amplitude, parallel propagating Alfv\'en waves subject to parametric instabilities. We have investigated how their stability, nonlinear evolution and saturation is affected by the amplitude of the pump wave, the plasma beta and the dimensionality of the system. Our main findings can be summarized as follows:
\begin{itemize}

\item In multi-dimensional systems the initial decay process can be described as a superposition of both the parametric decay instability and the filamentation/magnetosonic instability. The former leads to parallel propagating density fluctuations and Alfv\'enic modes, while the latter leads to the formation of perpendicular pressure-balanced fluctuations of density and magnetic field. The filamentation instability becomes the dominant process at beta values larger than unity, contrary to one-dimensional systems where the decay appears to be strongly suppressed. 

\item The decay process naturally results in a well developed turbulent cascade preferentially in the transverse direction to the mean magnetic field, and that shows similar properties for all the plasma beta values considered. At saturation of the instability, the magnetic energy spectrum displays a Kolmogorov-like inertial range at large scales. At sub-proton scales, a steepened magnetic field and a flattened electric field spectrum is observed, displaying power-law scalings which appear to be consistent with observations and simulations of plasma turbulence.  We have  quantified the nature of the fluctuations at sub-proton scales by means of spectral ratio analysis and we found a strong magnetic compressibility at sub-proton scales which is consistent with KAW linear theory, though a discrepancy with KAW theory is found in the excessive level of density fluctuations (possibly due to particle noise).

\item When the decay occurs, the saturated state is always characterized by a heated plasma displaying a persistent field-aligned beam localized at the Alfv\'en speed. The beam forms also when electrons are cold ($\beta_e=0$), pointing to the fact that it is the field-aligned Hall electric field to play the dominant role in accelerating the beam of particles. Such electric field is enhanced at the steepened edges of the pump wave and we argue that it mediates wave-particle interactions via both Landau and transit time damping. Landau and transit time damping are expected to become effective once the beam is formed, so that there is an increased number of resonating particles with the mean field-aligned  electric field and compressible fluctuations, respectively. 

\item The overall proton heating is predominantly a one-dimensional mechanism. We argue that wave-particle resonances in the mean-field aligned directions contribute to parallel heating.    
The perpendicular heating is instead attributed to a stochastic mechanism that works as an isotropization process of the distribution function via pitch-angle scattering.   A somewhat larger parallel temperature is observed in multi-dimensions than in 1D simulations, which may be explained in terms of the additional contribution due to protons interacting with turbulent fluctuations.  

\end{itemize}

In conclusion, the decay process acts as a trigger to  develop a turbulent cascade and to enhance wave-particle interactions, the latter resulting in a field-aligned beam and efficient plasma heating, reproducing in this way some features which are observed in the solar wind.  Moreover, we have shown that the decay process remains efficient also at large values of the plasma beta ($\beta>1$) which makes these results  relevant not only to space plasmas, but also to astrophysical environments where the plasma beta can reach values well above unity. It will be the subject of future work to investigate further the (nonlinear) wave-particle interactions leading to the beam formation and to the observed plasma heating to corroborate the scenario proposed here. It will also be important to extend our results to the expanding solar wind, and to investigate the evolution of a spectrum of Alfv\'en waves within the Expanding Box model by including a population of alpha particles \citep{Maneva_2015} to assess whether the instability is favored by the expansion or not,  what is the contribution of the decay process to solar wind heating and beam acceleration, and how minor ions react/modify the overall evolution.

\acknowledgements{This research was supported by NASA grant \#80NSS\-C18K1211.  We also acknowledge the Texas Advanced Computing Center (TACC) at The University of Texas at Austin for providing HPC resources that have contributed to the research results reported within this paper. URL: http://www.tacc.utexas.edu.}

\bibliography{resubmit}{}
\bibliographystyle{aasjournal}
\end{document}